\documentclass[a4paper,12pt]{article}
\usepackage{preamble}

\title{Expertise Elevates AI Usage: Experimental Evidence Comparing Laypeople and Professional Artists}   
 \author{Thomas F. Eisenmann*\textsuperscript{1}  \hspace{0.04cm}
 Andres Karjus*\textsuperscript{2,3} \hspace{0.05cm}
 Mar Canet Sola\textsuperscript{1,2,4}  \hspace{0.04cm}
 Levin Brinkmann\textsuperscript{1} \\
 Bramantyo Ibrahim Supriyatno\textsuperscript{1}  \hspace{0.05cm}  
 Iyad Rahwan\textsuperscript{1}\\   
 \textsuperscript{1}Center for Humans and Machines, Max Planck Institute for Human Development \\ 
 \textsuperscript{2}Tallinn University \hspace{0.04cm}
 \textsuperscript{3}Estonian Business School \hspace{0.04cm}
 \textsuperscript{4}Academy of Media Arts Cologne \\
 {\normalfont*}shared first authorship; corresponding authors: \\
 eisenmann@mpib-berlin.mpg.de; andres.karjus@tlu.ee
 }
\date{\vspace{-0.8cm}}

\begin{document}

\maketitle
\begin{abstract}\normalsize\noindent
   Generative AI's novel capacities raise questions about the 
   future role
   of human expertise: does AI level the playing field between professional artists and laypeople, or does expertise enhance AI use? Do 
   the cognitive skills experts make use of in analyzing and drawing visual art also transfer to using
   these new tools? This pre-registered study conducts experimental comparisons between 50 professional artists and a demographically matched sample of laypeople. Our interdisciplinary team developed two tasks
   involving image replication
   and creative image creation,
   assessing their copying accuracy and divergent thinking.
   We implemented a bespoke platform for the experiment, powered by a modern text-to-image AI. Results reveal artists produced more
   accurate copies and more divergent ideas
   than lay participants, highlighting
   a skill transfer
   of professional expertise --- even
   to
   the confined space of
   generative
   AI. We also explored how well an exemplary vision-capable large language model (GPT-4o) would fare: on par in copying and slightly better on average than artists in the creative task, although never above best humans. 
   These findings highlight the importance of integrating artistic skills with AI, suggesting a potential for collaborative synergy that could reshape creative industries and arts education.

   Keywords: Empirical Studies of User Behaviour; Artificial Intelligence and Expert Systems; Natural Language Interaction

\end{abstract}

\section{Introduction}

\begin{figure}[hbt]
	\noindent
	\includegraphics[width=\columnwidth]{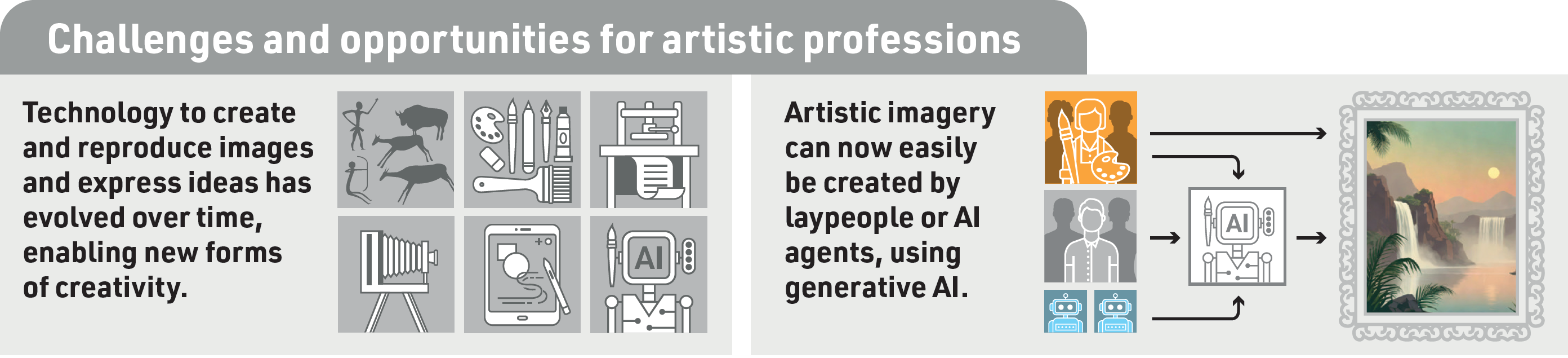} %
	\caption{Motivation for the study. 
    Left side: Generative AI can be compared to other technological developments that have shifted how visual art was made in the past, such as the industrial production of paint or the invention of photography. Right side: The traditional role of the expert directly producing art (top arrow) is now challenged by both laypeople and AI systems producing art via generative AI intermediaries. However, the skills of experts might also transfer to using generative AI themselves (second arrow from the top), opening up new opportunities as well.   
    }\label{figure_schematic}
\end{figure}

Generative machine learning models' increasing prevalence and capacities are transforming creative processes. Large language model-driven text generators and chatbots like ChatGPT or Copilot enable sophisticated text production. Generative image models and services such as Stable Diffusion, Dall-E or Midjourney enable the creation of artistic, photo-realistic, and illustrative visual materials without necessarily having professional training in these fields.
Reactions to the rapid adoption of these technologies have been strongly polarized. Alarmist responses range from claims that "art is dead" \parencite{roose_i-generated_2022} to reports of "AI anxiety" of workers fearing for their jobs \parencite{cox_ai_2023}, and questions about "creators becoming redundant" \parencite{dege_ai_2023}. A recent study found a significant drop in job postings for writing and image creation jobs on online freelance platforms after the introduction of chatbots like ChatGPT and image generation tools like Midjourney \parencite{demirci_who_2023}. Others compare the rise of generative AI to past technological advancements, such as photography and computer graphics \parencite{epstein_art_2023}. Viewing generative AI as a novel tool in the artist’s repertoire, they argue that these technologies do not replace art or the artist, but rather transform creative practices and give rise to new professions \parencite{epstein_art_2023},
in line with broader human-centered AI perspectives of AI as an augmentation rather than replacement of human capabilities \parencite{ozmen_garibay_six_2023}.
While all new technological developments can lead to a reassessment of traditional skills (e.g., \cite{arora_end_2013}), generative AI in particular might open up new pathways of art production (Figure \ref{figure_schematic}).
What will the future of creative work look like, and what will be the role of expertise in it?

The criteria for pinning down what exactly "makes an expert" differ widely, in art as well as in other domains. In a well-cited account, \textcite{shanteau_performance-based_2002} mention long-standing experience, external certification, social acclamation, consistency in judgments, and more possible characteristics used in the past. Ultimately, they dismiss most of these in favor of focusing on individuals' performance in their respective domain. This approach laid the groundwork for a science of expert performance \parencite{ericsson_capturing_2007}. In visual art specifically, psychological science has taken this performative view to heart and studied directly how artists' behavior and cognition differ from those of non-experts (see \cite{kozbelt_integrating_2007}, and \cite{ericsson_expertise_2018},  for summaries). Here, the focus is shifted from how to detect experts at all to detecting what makes experts psychologically special. Furthermore, this approach emphasizes the study of experts who are practicing art, as opposed to experts mostly judging it. In a formative study, \textcite{kozbelt_artists_2001} compared art students to students of different subjects in several tasks assessing skills in both drawing and visual perception. Perhaps unsurprisingly, art students performed better when copying images by drawing, but their superior results remained even when tasks required no motor execution, but only visuo-spatial cognition. Based on these results, the author argues that artists develop specialized cognition for analyzing the structural features of objects through their vast experience in planning and executing visual art.
This interpretation aligns well with the bulk of the literature comparing artists to non-experts. For instance, artists have been shown to exhibit better visual memory \parencite{obler_is_1988, perdreau_drawing_2015}. Work advancing the foundational study design has extended on some of the cognitive processes, showing that artists exhibit advantages in visual encoding \parencite{glazek_visual_2012, perdreau_drawing_2014}, visual selection \parencite{kozbelt_visual_2010, ostrofsky_perceptual_2012}, and local visual processing \parencite{chamberlain_local_2013, drake_artists_2024}. %
In summary, we can state that artists' expertise has been shown to be associated with quantifiable cognitive advantages. However, whether these advantages will translate to a new technology such as generative AI remains an open question. This is the driving primary research question being addressed here:
\textbf{
Can artists' skills transfer to using generative AI for image production?
}
Why should there be skill transfer from visual art expertise to using generative AI? We suggest that two criteria must be fulfilled to expect a reasonable advantage for artists. First, the task at hand has to be related to professional skills and practices. Some studies suggest that artists' skills transfer to visual tasks more general than the ones mentioned above, but not to logical reasoning, arithmetic, or lexical tasks \parencite{angelone_skill_2016, perdreau_drawing_2014}. This means that, trivially, we have to test artists on generative models for image creation, and not text or music models; but less trivially so, that the tasks administered also have to make use of visual cognitive abilities. We follow this logic by applying a copying task in the style of the psychological literature discussed earlier. In line with \textcite{kozbelt_artists_2001}, we interpret the classical drawing tasks as combined measures of both visual cognition and motor execution at the same time. In an equivalent task using the prompting of generative AI, we thus measure visual cognition and expression in language simultaneously. Since we are not leaving the domain of visual art, we expect that artists also excel at commanding relevant vocabulary and accurately describing objects, styles, and image structure.
Second, the change of medium has to leave room for skill expression at all. Pen-and-paper drawing has been studied heavily since it is seen as the natural example of visual artistry, and as minimizing the necessity of domain knowledge \parencite{ericsson_expertise_2018}. In theory, prompting as the predominant interface to generative AI models  could result in an even lower entry threshold.  With very capable models, laypeople might be able to engage in creative work at a similar level, without a need to learn how to draw. Alternatively, when using a bad generative model, everyone might simply be at the mercy of the machine translation, with no skill involved. There are reasons to doubt these accounts, however. In fact, instructing generative text and image models has been argued to be skillful, to the point of being called "prompt engineering". This is because informative inputs, which are well aligned with the pre-trained "expectations" of the model, tend to produce superior results \parencite{liu_design_2022,oppenlaender_prompting_2025, wei_chain--thought_2022}. Importantly, prompt content continues to matter even as models improve \parencite{jahani_as_2024}.  As such, we hypothesize that artists should be able to apply their existing visual skills also in this novel domain, beyond mere knowledge of generative models that they might also happen to possess.
In the classical drawing tasks, creativity is often explicitly discouraged \parencite{ericsson_expertise_2018}, since their aim is to get a clean measure of accuracy. Some research has focused not on the visuo-spatial processing of artists, but their creative skills instead; finding, e.g., that they produce more distant ideas in free association tasks than scientists \parencite{merseal_free_2025}. Defining creativity is a difficult endeavor, but a standard compromise that can be operationalized is to say it consists of \textit{original ideas that are useful} \parencite{runco_standard_2012}. Following the associative theory of creativity \parencite{mednick_associative_1962}, creative thinking requires making connections between seemingly remote concepts. This suggests that within a given domain, expertise could be helpful, as experts have a better knowledge and command of the specific concepts needed. Indeed, it has been argued that creativity requires a certain amount of expertise -- and that being creative is as domain-specific as being an expert in something \parencite{baer_importance_2015}. As such, the same questions about the skill transfer of experts become relevant also for creativity in using generative AI. 
The most common method to assess creative skills is via versions of the divergent thinking task, characterized by asking participants to produce original solutions to open-ended problems. Divergent thinking tasks should be seen more as measures of
creative potential than real-life creativity \parencite{Runco01012012}, but they have been shown to be predictive of creative achievement \parencite{reiter-palmon_scoring_2019, said-metwaly_divergent_2024}. There are diverse implementations of divergent production, as it has been shown to be domain-dependent \parencite{van_welzen_does_2024, vartanian_measurement_2019, zeng_can_2011}. As an example, a recent task specifically adapted for the creative production of music showed higher creativity in professional musicians, highlighted by more complex improvisations that are more distant from a prime melody \parencite{van_welzen_does_2024}. We follow this line of divergent thinking research by also testing a version of the copying task that explicitly instructs participants to diverge. Our hypothesis states that artists should also be able to express their strong creative skills in using generative AI.
Recently, the (more or less) independent production of art by AI itself has also gained increased attention. AI systems have shown proficiency both in the creation of content that would require high levels of craftsmanship \parencite{nightingale_ai-synthesized_2022, oppenlaender_creativity_2022, porter_ai-generated_2024,park_human_2024} and the exploration of novel conceptual ideas \parencite{bellemare-pepin_divergent_2024, haase_artificial_2023, hubert_current_2024, mirowski_robot_2024}.
A growing body of research is exploring the use of generative AI models, particularly large language models (LLMs), as autonomous agents across a broad spectrum of tasks, including customer service \parencite{wang_opendevin_2024},  software development \parencite{pandya_automating_2023}, publishing books \parencite{guljajeva_synthetic_2022}, and artistic production \parencite{klingemann_botto_2022,guljajeva_ena_2021}. Many of these applications demand forms of creativity, such as divergent thinking. The empirical evidence on whether generative AI can perform creative tasks at a level comparable to humans remains inconclusive.
On the one hand, it has been shown that current-generation LLMs are not only theoretically capable of reasoning and creativity \parencite{wang_can_2024}, but already outperforming humans in various creative writing, persuasion, and divergent thinking exercises \parencite{argyle_leveraging_2023,bellemare-pepin_divergent_2024,hubert_current_2024,mirowski_robot_2024,wu_one_2025}, 
formal writing and research tasks \parencite{lehr_chatgpt_2024, schmidgall_agent_2025, wen-yi_automate_2024},
as well as various forms of text annotation and analysis \parencite{karjus_machine-assisted_2025, gilardi_chatgpt_2023, tornberg_chatgpt-4_2023, ziems_can_2023}.
Being trained on human-produced data, they also exhibit similar biases \parencite{acerbi_large_2023,kotek_gender_2023,motoki_more_2023}.
On the other hand, some work has reported humans as being more imaginative than LLMs \parencite{begus_experimental_2024}, sometimes depending on the task \parencite{charness_creativity_2024}. Another study found LLMs being more creative, but less diverse  \parencite{doshi_generative_2024}. Naturally, all such comparisons depend on which humans the machines are compared to, the average or the top performers, experts or laypeople \parencite{koivisto_best_2023,haase_artificial_2023,porter_ai-generated_2024}. This leads us to 
our second, exploratory question: 
\textbf{
Can an AI system match the results of artists?
}
\subsection{The Present Research}

To address the questions outlined above, we need to empirically investigate the relevant population: practicing professional artists. We present a unique sample comprising 50 professional artists and 49 matched laypeople, and evaluate their performance in interactions with a text-to-image generative AI system. Additionally, we assess the performance of an example LLM, acting as a prompt writing agent, on the same tasks. In doing so, we provide further insight into the current capabilities and limitations of models that can be used in autonomous creative systems.

We test a set of preregistered hypotheses 
concerning our expectations that artistic skills should transfer to 
using generative AI, based on the literature and reasoning outlined above.
We devised two controlled experimental tasks, emulating activities common in both artistic practice and education: replication or copying of existing art, and the creative production of novel or divergent art (see Figure \ref{figure_design} and Procedure for details).
These are inspired by traditional copying and divergent thinking tasks, respectively. In line with that research, we use distance measures to evaluate both tasks, suggesting that smaller distance is better in the copying task, but higher distance is better in the creative task (see Measures).

Generative AI usage typically combines both creation and curation, as it is easy---and often the default workflow in many applications---to produce concurrent variants based on a single input, and choose the most suitable output. We therefore enable and analyze curation and selection by participants.
We expect artists to excel also at curation, based on the same advantages in visual analysis: They effectively get another chance at visual comparison. We acknowledge this with similar hypotheses also for the set of curated images, plus an extra hypothesis that considers curation independently of the skills used in the creation of images.
This structure is reflected in our concrete hypotheses:

\textbf{\textit{General performance:}}
\begin{itemize}[leftmargin=0pt, itemsep=0pt, topsep=0pt]
 \item[] H1: Artists' images in the copying task will overall be closer to the original images than laypeople's.
 \item[] H2: Artists' images in the creative task will overall be more distant from the original images than laypeople's.
 \end{itemize}
 
\noindent
\textbf{\textit{Curated performance:}}
\begin{itemize}[leftmargin=0pt, itemsep=0pt, topsep=0pt]
 \item[] H3: Artists' curated images in the copying task will be closer to the original images than laypeople's.
 \item[] H4: Artists' curated images in the creative task will be more distant from the original images than laypeople's.
 \item[] H5: Artists will consistently choose more suitable images from the selection of four images in the pooled data of both tasks, compared to laypeople.
\end{itemize}

\noindent
 \textbf{\textit{Explorative part}} \\
We registered that we explore a comparison of the human groups to the outputs of a similarly instructed AI language model in an approach comparable to H1-2, but did not commit to a prediction about its performance.

\section{Method}

\subsection{Participants}

Our final sample consists of 99 participants: 50 professional visual artists, and a matched control group of 49 laypeople recruited via the crowdsourcing platform Prolific. The artist sample was recruited first, via invitation emails sent directly through the coauthors' 
(mostly MCS)
 personal and professional networks. This way, we ensured artists fulfilled our pre-registered recruitment criteria (see below) and avoided excessive over-recruiting and exclusion. After the full artist sample had been completed, we recruited the laypeople sample to match its demographics. 

All participants were paid €3.75 after completing the full experiment, in compliance with the ethical approval received for the study beforehand (by the IRB of the 
 Max Planck Institute for Human Development,
approval number A2024-15). Some artists (22\%) were more motivated by participating in a scientific experiment than by the remuneration itself and waived their payment.

There were a number of criteria that participants needed to fulfill, or otherwise they were immediately excluded during recruiting and replaced with a new random participant.
All participants were required to be fluent in English, but were allowed to have a different first language. Most importantly, the artist sample had to actually represent professionally working artists, whereas the laypeople sample must not, to ensure a clean separation of the two conditions. We made sure of this by recruiting the artists from our network of professional artists, and double-checking that the Prolific sample did not include any. There was an item in the post-questionnaire to screen for participants' correct assignment to their condition ("Do you have work experience in visual art?"). 
If a participant's answer did not match their condition, they were excluded from the sample. 
This happened in 5 cases for artists (e.g. having art education but not working professionally as an artist) and in 10 cases for laypeople (due to chance, because we could not initially screen them against that).

We also inquired about the number of years the artists had been active in visual arts, which was high in our final sample ($M = 22.4$, $\sigma = 10.7$). 
We chose to focus on artists in professional careers, defined as those actively engaged in creating art and participating in the distribution system, such as exhibiting or selling their work. This has been called the most salient marker distinguishing serious artists from amateurs \parencite{becker_art_1982}. Thus, art students and academics not actively working in the field were excluded. We further narrowed our focus to \emph{visual} artists, as this was the medium of the task.

While recruiting the artist sample, we recorded participants' highest level of education and their first language, identically to two screening items on Prolific. We then used the screening function on Prolific to match the participants faithfully, according to their specific combinations of education and language (binary English/not English). As expected, artists' highest level of education was very high, with 24 participants of the 50 stating they had a Master's degree and 23 more reporting a PhD. 14 artists reported being native speakers of English. Additionally, we use self-reported AI usage experience levels as a control variable in the results section; here, artist participants were more experienced ($M = 1.68$) than laypeople (1.04, on a scale of 0 to 3).
Apart from group assignment, we specified two more exclusion criteria in the preregistration. Participants who did not enter a prompt in two or more trials were replaced due to too much missing data. Images that were blanked out for participants due to our implemented filter (see below) were counted toward this limit. These criteria led to the exclusion of two more participants (one per subsample).

\subsection{Procedure}
The current experimental design draws from several experimental traditions. These include interactive behavioral experiments often used in cognitive science and adjacent disciplines \parencite{kirby_cumulative_2008,okada_imitation_2017,nolle_emergence_2018,muller_influence_2019,karjus_conceptual_2021,hawkins_visual_2023,kim_collecting_2024}, 
research comparing the behavior or outcomes of domain professionals or experts with some control group of laypeople \parencite{kozbelt_artists_2001,bhattacharya_drawing_2005,bezruczko_differences_1994,torngren_worse_2004}, 
and studies on individual aesthetic perception and preferences
\parencite{porter_ai-generated_2024,cela-conde_sex-related_2009,lakhal_beauty_2020}.

\begin{figure}[hbt]
	\noindent
	\includegraphics[width=\columnwidth]{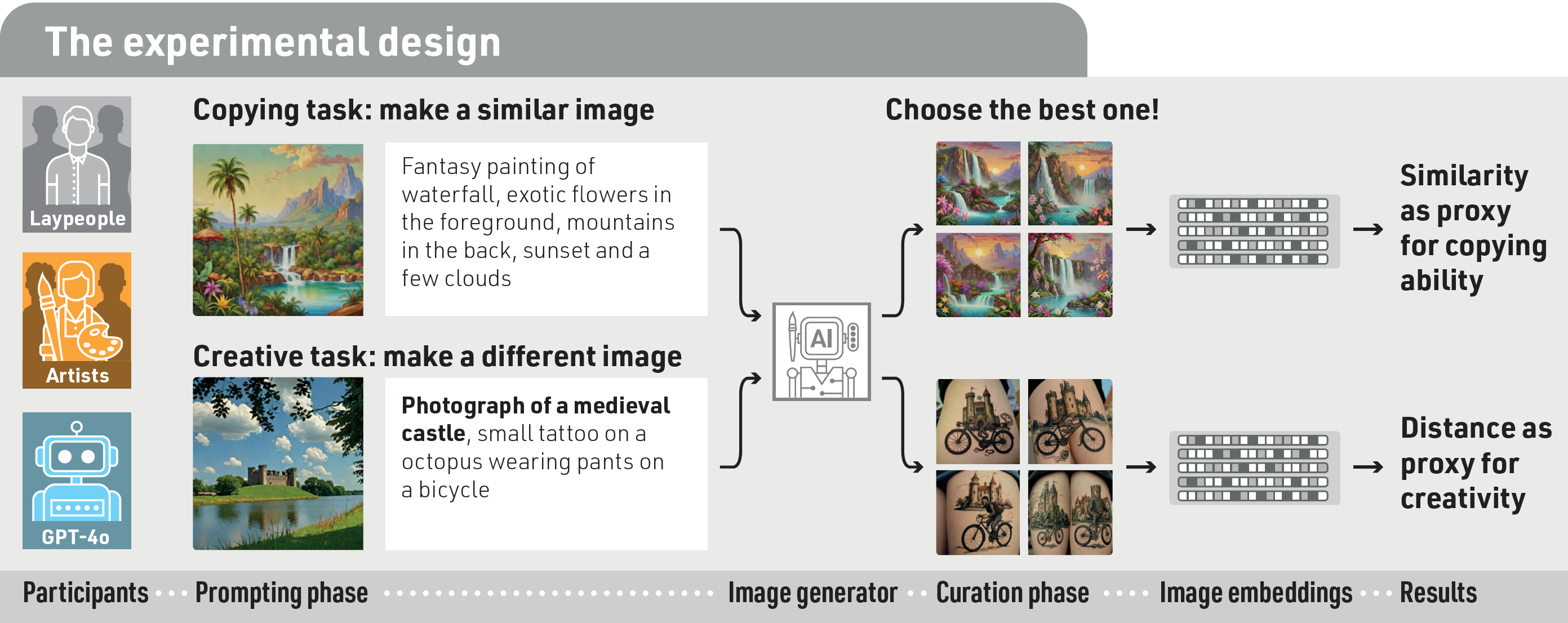}
	\caption{
    Pipeline overview. The participants were asked to view a reference image and write a prompt for a generative model to either create a similar image (copying task) or a maximally different image (creative task). In the curation phase, they were shown four generated variants and asked to select the most suitable one given the goal of the task. We later compared the similarity of the reference and generated images using an image embedding model, and used cosine vector similarity to operationalize the results. The prompts and variants are from one of the artist participants; the rest of the images created by this participant can be found in the 
    Appendix.
    }\label{figure_design}
\end{figure}

The online experiment consisted of two distinct tasks,
which we call "copying" and "creative" for short. The tasks were divided into "prompting" and "curation" phases, and each task contained four such image generation and curation trials.
Figure \ref{figure_design} illustrates the pipeline from experimental tasks to results.

In both tasks, participants were
shown a reference image and asked to write short textual inputs ("prompts") to produce new images and select their preferred output.
We created a bespoke online platform for the experiment, aimed to broadly mimic currently common generative AI apps and platforms. 
The experiment started with a page where participants could give informed consent to the study by choosing that option within a written form. Then they read the study instructions, followed by two training trials where the participants were free to generate images from their own prompts to familiarize themselves with our generator, the interface, and the prompting and curation phases. They were then instructed in the first task, completed it, got instructed in the second task, and finally were asked to complete a short questionnaire. In total, the average participant took about 18 minutes to complete the experiment.

Participants completed the copying task first and creative task second. There were 4 trials within each task which displayed a unique reference image. The order of images was randomized within each task. Each of the 8 trials consisted of a "prompting phase" and a "curation phase". The interfaces were the same for both tasks.
In the prompting phase, participants were tasked with writing prompts to create images from. Here, the interface showed the unique reference image, a text entry box, and a "submit" button. Prompt length per trial was limited to 30 (space-separated) words, visible as a counter below the text box. Furthermore, there was a time limit of 2 minutes per attempt, which was also displayed. If time ran out before participants had submitted their prompt, their current input was submitted and they moved to the curation phase automatically.
We chose these specific time and word limits to keep the task short and maintain comparability between answers, as they avoid excessive replies and ensure participants have to approach the task in a more standardized way.

After prompting, our generative model created 4 different images for the participant to choose from in the curation phase. The reasoning behind this was threefold. First, this mimics real-world generative AI usage, where tools often provide multiple variants by default. It also balances the randomness inherent in different seeds (see below). Finally, this enables testing whether curating would also allow for the expression of artistic expertise, as per our second set of hypotheses. The interface showed the 4 generated images, with the reference image displayed for comparison. Here, the time limit was 30 seconds, as participants only needed to click their preferred image to submit.

In the copying task, participants were asked to prompt the model to generate an image "as close as possible" to the reference image. 
In the creative task, participants were instead asked to make a new image that would be "as different as possible". Their instructions further clarified that 1) ideally, they should aim for "an image with different \textbf{content} as well as different \textbf{visuals}", and that 2) negation does not work well for this and would actually increase the likelihood of adding something to the image. To make sure results remain comparable and the task reasonably challenging, the start of the prompt was (visibly) fixed in the text box to the first words used to create the reference image. For example, if the prompt we used to create the image was "A photo of a classic gray Mini Cooper in a parking lot of a shopping mall", the text box would contain the locked starting phrase "A photo of a gray classic Mini Cooper". %

\subsection{Material}

The text-to-image model in the backend was Stable Diffusion XL Turbo (henceforth SD), chosen for its ability to generate images within seconds 
and its open source nature 
\parencite{sauer_adversarial_2023}. 
The outputs of SD and similar models are inherently affected by a random seed number.
To account for this randomness, we used four fixed seeds to generate the variants that participants curated from. These seeds were distinct from the seeds used to create the reference images, so participants could not reproduce the reference images even if they found the exact corresponding prompt.
This also makes the outputs comparable: as all parameters are fixed for all users, if two participants entered the exact same prompt, they would get the exact same set of four images. 
A keyword-based filter was used to prevent users from creating sensitive or obscene images, based upon techniques recommended by 
\textcite{rando_red-teaming_2022}.
 This was for ethical concerns and because creating such images was not instrumental to the task. Participants were informed about the filter and asked to avoid creating inappropriate content.

The eight reference images were created using the same SD model by our professional artist coauthor (%
MCS) to ensure sufficient quality and task difficulty. We aimed to cover a diverse range of artistic imagery, subjects, and styles: landscapes, architecture, vehicles, food, humans, and cartoon characters; and photography, animation, painting, drawing, and video game 3D styles. Additionally, we kept the original prompts in the copying task relatively complex and the ones in the creative task relatively plain; this was to avoid ceiling or floor effects. 

We have published all the image data resulting from the experiments (see Data Availability) and also produced an interactive dashboard based on the Collection Space Navigator \parencite{ohm_collection_2023} that enables easy exploration of the dataset, available here:
 \url{https://artistlaypeopleaiexperiment.github.io}.

\subsection{Measures}

\subsubsection{Dependent Variables} 

As preregistered, we used the CLIP \mbox{(clip-vit-base-patch16)} embeddings \parencite{radford_learning_2021} to quantify the similarity between reference images and participant creations, via cosine similarity of the embedded image vectors.
CLIP translates images and text into numerical representations (“embeddings”) in a shared multidimensional space. In this space, images that are visually or conceptually similar are located closer together, whereas dissimilar images are farther apart. Measuring the cosine similarity between two embeddings therefore provides an estimate of how similar their visual or semantic content is.
As outlined below, this way we measure the constructs of Copying Accuracy and Divergent Thinking in the copying and creative task, respectively.
SD models are known to use CLIP internally for encoding prompts, making this a natural choice. Our automatic assessment has the advantage of being flexible enough with regard to the complex imagery produced in the task, while remaining more objective, scalable, and reproducible, compared to traditional subjective ratings by humans. For comparing groups in H1 and H2, the values were averaged for each set of four variant images, as we are more interested in prompt outcomes rather than individual images.
\textbf{Copying Accuracy.}
In the copying task, the goal was to create a close replication of the reference image --- therefore, higher cosine similarity indicates a better match. 
A distance-based measure is in line with the approach used in traditional copying tasks, based on either subjective raters (e.g. \cite{kozbelt_artists_2001}) or objective measures based on geometry (e.g. \cite{carson_angle-drawing_2013}). Importantly, Copying Accuracy in our task represents the outcome of two separate cognitive processes: the visual analysis required to process the reference image, and the expression of that image via a natural-language prompt. While the advantages of artists in visual cognition have been well-established (see Introduction), closer resulting images would indicate that these skills transfer also to generative AI through language expression; i.e., that the change of medium does not eliminate the cognitive advantages, but enables their expression.
\textbf{Divergent Thinking.} In the creative task, where the goal was to create a different image, lower cosine similarity indicates a better outcome. 
This is in line with a recent methodological shift to assess divergent thinking tasks via semantic distance (e.g. \cite{koivisto_best_2023}; \cite{yu_mad_2025}). In an in-depth evaluation, semantic distance was shown to strongly predict creative performance on a range of tasks \parencite{beaty_automating_2021}. Conceptually, higher distance is an indicator for the remoteness of the implemented idea. This remoteness is based on artists' visual cognitive skills, since creativity depends on expertise \parencite{baer_importance_2015}. However, as in the copying task, distances are also the results of language expression in prompting. This means that more distant resulting images could be due to more remote ideas, better expression of ideas in language, or both. This is somewhat mirroring the concept that creative results should be both original and useful. Crucially, a positive result for artists still would suggest that generative AI allows for the transfer of skills also within a creative context, no matter the exact causal relationship. 

We wanted to make sure the 
divergent thinking measure could not be manipulated
by simply filling the prompt with random words or random strings, which might conceivably lead to images different from the reference. We ran a small simulation to test that by producing a set of prompts of maximal length out of random words sampled from an English word list, and another set consisting of pseudo-words of randomly sampled letters, and generating new images based on those. We found that while it is possible to occasionally "get lucky" with this strategy, adding random things to a prefixed prompt generally leads to images still close to the reference image, and the simulated results were worse on average than the real data (see SI for details).

\subsubsection{Control Variables}
\textbf{Highest Level of Education.} Artist participants were asked "Which of these is the highest level of education you have completed?", choosing from "no formal qualifications" to "doctorate degree". This was the first of two variables used to match laypeople later in the recruiting by using an identical screening item. Education was chosen to make sure any differences found in the tasks would be due to domain expertise and not general academic background, since we expected (and found) that artists were, overall, highly educated.
\textbf{First Language.} Artists were also asked "What is your first language?", in line with the matching item later used in the laypeople sample. This matching was used to control for language abilities that could mask or increase the differences between artists and non-artists, given that prompting is English-language-based and the whole task was presented in English. Consequently, we matched in a binary fashion of "first language English" or not.
\textbf{AI Experience.} We used an item in the participant survey to statistically control for usage experience with generative AI in our analyses. Participants were first asked "Did you ever write prompts for generative AI (e.g. Midjourney, ChatGPT, DALL-E, etc.) before you started the experiment?" to clarify what we intended to measure. Then, we recorded their experience ("Please rate your experience with prompting for generative AI before you started the experiment.") from "none" (if they said "no" initially) to "expert" on a four-point scale. The purpose of this measure was to separate prompt engineering, i.e. specific skill with generative models, from the visual-cognitive skills we are interested in.

\FloatBarrier

\subsection{AI Procedure}

Inspired by the growing literature on comparing humans and various AI agents discussed in the Introduction, we also carried out a small comparison with the vision-capable large language model GPT-4o by OpenAI (specifically \mbox{gpt-4o-2024-08-06}, via its API service). The LLM was given roughly the same instructions as the human participants in the experiments, with added context that was otherwise implicit in the task interface (such as the word limit), and general guidance to act in the role of a creative professional.
It was instructed to write a prompt of up to 30 words just like human participants. Any words above the limit would be clipped to ensure comparability with the rest of the experiment (there were only 3 such occasions, however).
In short, we created something akin to a simple image evaluation and generation agent, consisting of three models: the LLM interpreting the input and producing the prompt, the image generating SD, and the CLIP model yielding embeddings for our goal of image comparison.

The procedure was as follows: for each trial, the input was the reference image and the instructions to carry out the given task, e.g., to write a prompt that would generate an image like the reference forest image. The model was set to generate 10 output variations on each trial. Unlike participants, the LLM here was not granted a "memory" of past completed attempts. The prompts were subsequently entered into the same image generator used in the experiment, with the same parameters, to produce images that could then each be compared to the reference, as described above. 

\section{Results}
\subsection{Data Exclusions}

After data collection of all 100 participants had concluded, it became apparent that the prompting data of a single participant from the laypeople group had not been passed on due to a technical error (while the participant provided all other information), and was thus unusable; hence we arrived at our final sample of 99 participants. After inspecting the data, we identified and removed four trials that we deemed ineligible. These included errors like a single-letter prompt, two cases where participants seemed to have initially misunderstood the task, and one case where the fixed lead of a creative trial merged with the input due to a bug.
We came across 11 occasions where participants had circumvented the whitespace-operationalized word limit by concatenating a few words at the end of the prompt, but did not deem this an exclusion criterion. While clever, omitting spaces carries the risk of confusing the text tokenizer of the model and creating unexpected visual outputs. The resulting final dataset consists of 3148 images. 

\begin{figure}[htb]
	\noindent
	\includegraphics[width=\columnwidth]{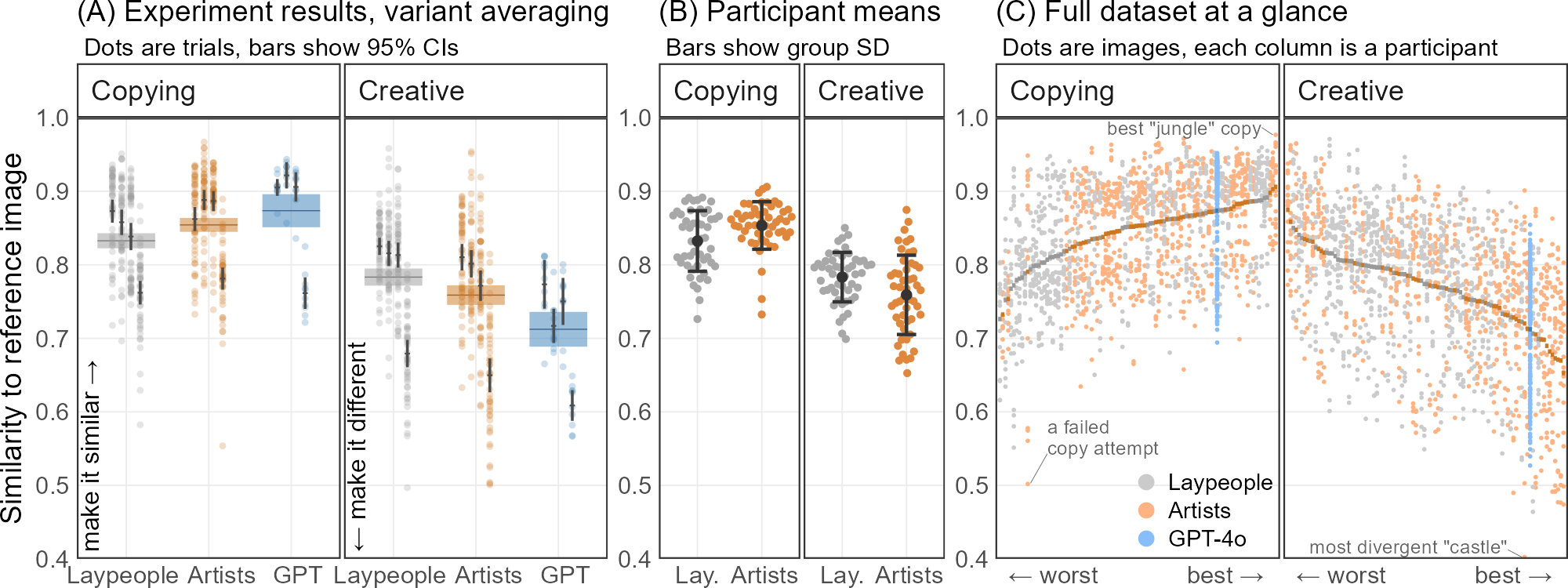}
	\caption{
           Experimental results: artists are better at using generative AI than laypeople. The vertical axis is cosine similarity; for the copying task, the goal was to produce similar images (high values), and for creative, dissimilar. The panels display three views of the data. Each dot in (A) represents the distance of a generated image from the reference (averaged across its four variants), the central unit of analysis. The columns of dots are organized by the four stimuli in the two tasks. Some images were harder than others, e.g. the "couple" image was most difficult to copy for all groups (rightmost dot column in Copying panel; cf. images in Figure \ref{figure_examplegrid}; see the SI for an extended comparison graph). The darker error bars show stimuli means with 95\% confidence intervals. The three colored thick bars are the group averages in both panels, with the shaded area showing the CI. We confirmed group differences using mixed-effects models that control for stimulus variance, participant variance, and also for participant's prior AI experience. 
           Panel (B) displays participant-level data, variant means averaged for each participant; standard deviations of these averages by group are shown as black bars. There is more participant variance in the creative task, and more so among artists than laypeople.
           Panel (C) displays the entire dataset, arranged by participants (dot columns) according to their mean task performance (line of rectangles). GPT-4o does not produce the best works but does better on average than laypeople in the copying and both groups in the creative task.  
    }\label{figure_comparisons}
\end{figure}

\subsection{General Performance of Artists and Laypeople (H1 and H2)}

\begin{figure}
	\noindent
	\includegraphics[width=\columnwidth]{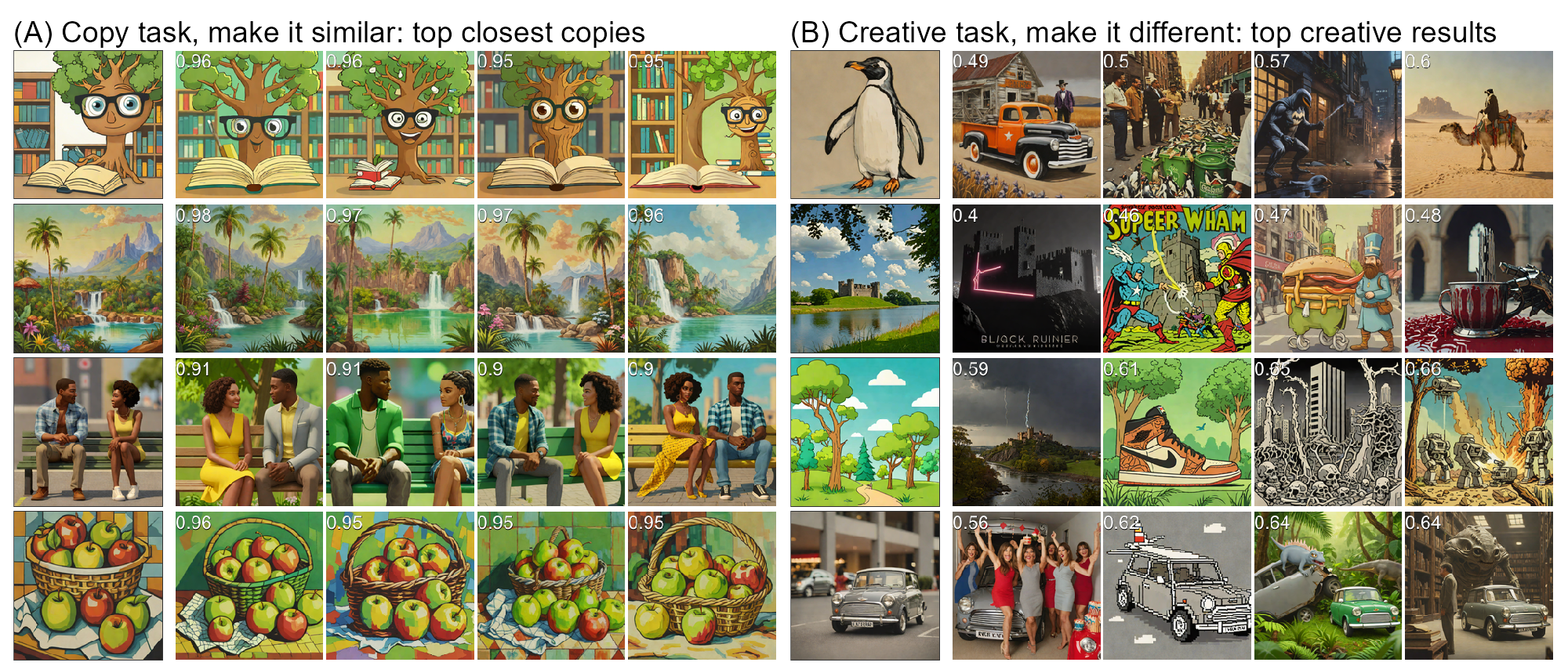}
	\caption{Examples of most successful trials across the experiment; cosine similarity in the corners. Panel (A): closest copies for each reference image (left, black border). (B): the most successful creative results, diverging furthest from the reference, either by managing to shift the style, hide the prefixed subject, or transform it. A larger version of this graph, comparing the best to the worst results, can be found in the 
    Appendix.
    }\label{figure_examplegrid}
\end{figure}

Figure \ref{figure_comparisons} illustrates the results of the experiments.
We use mixed-effects generalized linear regression to test H1-2 concerning the outputs of our two groups, in the following form:  \mbox{cosine $\sim$ group + experience + item + (1|subject)}. For both the copying and creative task, the dependent variable is the cosine similarity between the original and generated image. The main variable of interest is the group: the coefficient value of this shows how much the groups differ in terms of outcomes. We also control for participant background by fitting a fixed effect for (self-reported, numeric) prior AI usage experience, and fit random intercepts for participants to account for individual-level variation. Stimulus variance was also controlled for via a fixed effect --- given its small number of levels, there was no practical difference to a random effect, which yielded identical results. As discussed above, we use the averaged distance of the four generated variants as the unit of data here, yielding 787 cases. The reference images of both tasks and the top results are visualized in Figure \ref{figure_examplegrid}.

We find support for both hypotheses 1 and 2. In the copying task, the estimated cosine similarity of laypeople is -0.03 (95\% confidence interval between $[-0.04, -0.01]$) lower than artists, the baseline category of the group variable. The coefficient confidence intervals were estimated using bootstrapping with 1000 model replicates. The effect is significant (at $\alpha=0.05$), as determined via a likelihood ratio test comparing the full model to a partial one without the group variable ($\chi^2 = 11.1, p < 0.001$). The model intercept (average artist, zero experience, baseline stimuli category) is at 0.9 on the cosine scale, which theoretically ranges between $[-1,1]$, where 1 would indicate an identical embedding vector and therefore image. Here the maximal result is 0.98, the first jungle image in Figure \ref{figure_examplegrid}A. The model likelihood-ratio based \textcite{cox_analysis_1989} based pseudo-$R^2$ is 0.03, i.e. the group variable describes about 3\% more variance than the reduced model.

In the creative task, laypeople are 0.02 $[0.002, 0.04]$ \emph{closer} to the reference than artists ($\chi^2 = 4.56, p = 0.03$; model intercept 0.82, pseudo-$R^2$ is 0.01). Here the goal is to be distant, so the artists' result is superior. While statistically significant, these are not very large differences, as also visible in Figure \ref{figure_comparisons}A.
In both tasks, the best (closest or most divergent) individual images were generated by artists.
The cosine similarities of the images by artists have lower dispersion ($\sigma = .032$) than laypeople ($\sigma = .041$) in the copying task, but higher dispersion ($\sigma = .054$) than laypeople ($\sigma = .034$) in the creative task, as also illustrated in Figure \ref{figure_comparisons}B.
The effect of the control variable of prior AI experience was small, and its bootstrapped confidence interval included zero, indicating no reliable association.

\subsection{Curated Performance of Artists and Laypeople (H3-H5)}

The experiment combined both artistic image creation and curation, as participants were presented with four variant images to choose from after every generation trial. We can therefore assess the effect of curation as well. We constructed two sets of models to do so. Here we excluded the few trials that produced NSFW content, as the participants were not given a chance to curate those (blank images were displayed), leaving 758 cases. First, we ran the same mixed-effects linear models as above on this dataset reflecting curation choices. For both tasks, it resulted in similar differences: laypeople were more distant than artists in copying ($-0.03 [-0.04, -0.01], p=0.001$) and more similar in the creative task ($0.026 [0.003, 0.05], p=0.02]$), supporting H3 and H4.

To test H5, we also ran a cumulative link mixed model (with logit link function), again also controlling for prior experience and using random effect for subjects and stimuli. The dependent is a 4-level ordinal variable, reflecting the rank of the participant's choice in terms of the best option among the four, as measured by CLIP similarity (where 1 is the furthest, 4 is the closest; for the creative results this is flipped to be comparable). The model indicates no difference between the groups when it comes to choosing between image alternatives ($p=0.71$, bootstrapped coefficient confidence intervals span 0). However, most variant image sets were already highly similar (average cosine similarity at 0.93 in the copying and 0.91 in the creative task), effectively leaving little room for potential differences in curation skills to be expressed.

\subsection{Exploration of AI Performance}
We also explore an additional comparison to GPT-4o, one of the frontier multimodal LLMs currently also powering the popular ChatGPT chatbot service. The statistical modeling solution here is less complex, as there is only one "participant" in the GPT group, and there is no comparable experience variable for a machine.
We therefore fit a simple fixed-effects linear regression model, where the group variable now contains the GPT, set as the reference level (we fitted a random effects model as well which yielded very similar results). As in the first comparison, we average the variant image cosine similarities for each trial. In the copying task, there is no significant difference between GPT and artists ($p=0.11$). Laypeople are an estimated -0.04 further from the reference ($\mathrm{CI} =[-0.06, -0.02], p<0.001$; model adjusted $R^2 = 0.03$), meaning GPT did better here. 
In creative, where the goal is to have a lower value, both laypeople ($\beta=0.07 [0.04, 0.1], p<0.001$) and artists ($\beta=0.05 [0.02, 0.08], p=0.002$; model $R^2=0.05$) are on average higher than GPT. This means GPT-4o was able to write prompts that led to on average more creative visual results than both human groups, within the narrow definition of the task. The variance described is very low in both models, however, less than 5\%, corresponding to the rather small absolute differences, and the top results in each task are still achieved by the most successful human participants, mirroring results by \textcite{koivisto_best_2023}.

\subsection{Exploration of Image Diversity}
We explored the diversity of images in the creative task by first measuring variation in the images and then fitting a linear mixed-effects model. Variation is implemented as the cosine similarity of each image embedding vector to its group centroid (average) vector, computed for each trial image and group. This is analogous to mean absolute deviation or MAD. The regression then measures the effect of the group on the dependent variable of the variation metric (with random effects for group and trial reference image). The unit is still the cosine (same scale as in Figure \ref{figure_comparisons}), so higher values here indicate closer proximity to the center, i.e. lower variability. Laypeople ($\beta=0.02, \mathrm{CI}=[0.003, 0.03]$) and GPT-4o ($\beta=0.05, [0.001, 0.11]$) have both higher cosine, i.e. lower collective diversity than the reference level of artists (model $p=0.01$ compared to reduced model). 

As a second approach, we also measured diversity within participants: how much the four images (and their variants) differed from each other on average. In the creative task, some participants (and GPT) used similar strategies for all four trials, while others varied their approach (see example image sets in the SI). We measure this as the average of all pairwise embedding similarities within a given participant. For GPT, we simulate subjects by replicating the dataset 10 times, randomly recombining the 10 image prompt-variant sets (keeping the output images of a prompt intact). Here, the statistical model is simple regression, as trials and participants are already averaged. The artists have an estimated average intra-similarity of 0.63 (still on the same cosine scale), and laypeople do not significantly differ ($p=0.91$), while GPT yields higher estimated similarity, i.e. less diversity than artists ($\beta=0.07, p<0.001$, model $F(2, 196)=155.8, p<0.001$).

\subsection{Exploration of Prompt Content}

We also carried out an explorative analysis of prompt components, using a multi-factorial feature analysis or quantitizing design. As abundantly demonstrated in recent literature, modern LLMs can be utilized as convenient on-demand classifiers and information retrieval engines in lieu of traditional NLP pipelines or human annotators \parencite[][]{karjus_machine-assisted_2025,ziems_can_2023,rathje_gpt_2024}. We use GPT-4o in a zero-shot manner with instructions to retrieve the following components, if present, from each of the 787 prompts:
artistic style or style period, genre (e.g. landscape, still life), medium (photo, painting, etc.), mentions of any color, presence of describing adjectives, mentions of subjects and tangible objects, and in contrast, just setting or background descriptions (jungle, library, night-time). We selectively reviewed the LLM outputs and found they generally matched the data very well.

\begin{figure}[htb]
	\noindent
	\includegraphics[width=\columnwidth]{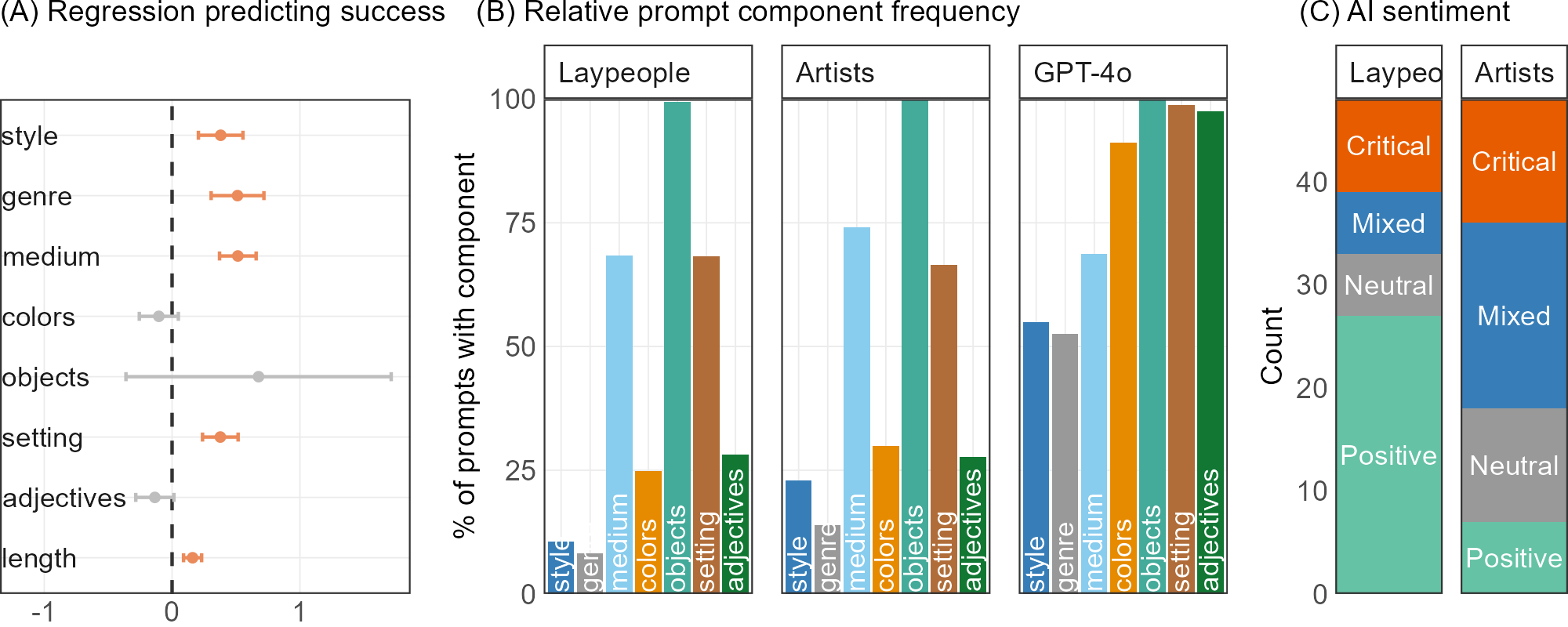}
	\caption{How do people write prompts? (A) Coefficients with 95\% confidence intervals, from a linear regression model predicting the (z-scored cosine) success score: higher values indicate better result. Orange color indicates significant effects ($p<0.05$). All variables except length are categorical, with no presence as the reference level (presence predicts the x-axis value of increase in standard deviations in score). Prompt length in characters is scaled separately for the tasks; a $1\sigma$ 
    increase in length predicts a $0.18\sigma$ increase in score. Panel (B) shows the relative usage frequency of the same components by the two groups, and (C) shows the sentiments towards AI reported by our experiment participants.
    }\label{figure_promptnlp}
\end{figure}

Here the two task datasets from human participants are concatenated for a joint analysis. They are made comparable by inverting the cosine similarities of the creative task and z-scoring (centering and dividing by standard deviation) these values separately for the two tasks. Prompt length, in characters, is similarly normalized by z-scoring separately, as creative task prompts had a fixed prefix the participants could not change and is not counted towards length. The prefix was removed from the feature analysis, and the creative prompts automatically coded as having an object and medium, as both were present in all prefixes.
Figure \ref{figure_promptnlp}A illustrates the results of a linear regression model predicting simple additive contributions of the (binomially-coded) presence of the components and prompt length to task success. Longer and more informative prompts do slightly better as expected. Specifying stylistic aspects and setting a scene helps, but concrete descriptions of subjects (along with adjectives and colors) are a double edged sword, as mis-specifying them can easily lead to suboptimal results. Figure \ref{figure_promptnlp}B illustrates the relative usage of these components between groups, which are surprisingly not that different, but artists do appear more likely to define the style, which their training would be expected to support.

\subsection{Exploration of Sentiment Towards AI}

We also analyzed the participant sentiment towards AI, elicited in the post-experiment questionnaire. Most left some comments, either on generative AI in general or on the generative tool used in the experiment (96 comments could be used, as a few were empty or illegible). As above, we used GPT-4o as a zero-shot stance detection tool, requesting the following categories: positive (including excited or enthusiastic), critical or concerned, mixed (multiple or ambivalent feelings expressed), and neutral or indifferent (including discussing pragmatic technical aspects). Figure \ref{figure_promptnlp}C shows that differences do emerge: 27 of the laypeople are positively minded about AI compared to only 9 artists, who were more likely to have mixed or critical feelings. Naturally, as a survey, this is a small sample, and such attitudes have been surveyed more extensively elsewhere \parencite{novozhilova_looking_2024,goetze_ai_2024,lovato_foregrounding_2024,tang_exploring_2024,lima_public_2025}, but it serves to provide some insight into our sample.

\FloatBarrier

\section{Discussion} 

The results indicate that expertise in visual art does provide an advantage in using generative AI for image generation. Artist participants created more 
accurate
copies and more 
divergent ideas, even under the severe time and technological constraints of the experiment, limited to a single generative model in a restricted manner. While generative AI has made content creation accessible and fast, the visual and creative cognitive skills of professionals remain relevant in effectively using these tools as well, painting a more nuanced picture of these evolving practices.
We see some qualitative evidence for this also within the content of artists' prompts, which were longer and contained more references to the style, genre, or medium of the image; all of which also contributed to better-performing results. At the same time, we can exclude that our main results were merely due to differences between the two samples in prompting experience, education level, or fluency in English, as these were controlled and matched for. Unexpectedly, we did not find differences in image curation between the two samples that went beyond the fact that the average image created by artists was already performing better due to better prompting. We believe that this is due to the variance between the 4 created images being very low, leaving little opportunity for the artist group to demonstrate their skills. However, this is a useful insight by itself, suggesting that seed randomness is not a threat to our results.

By design, our participant groups were not random samples. The artists were carefully selected in hopes to be representative of the artistic profession. The lay participants were intentionally matched to the education level and first language of the artists. 
Our recruitment procedure represents an attempt to balance control, feasibility, and ecological validity. An example of this is that several professional artists chose to forgo payment entirely, showcasing potentially different motivations behind participation in the task. While it could be argued that this might have introduced a differential in economic motivations for completing the task, we would reply that intrinsic motivation is part of what makes the professional artists artists, and that it may be impossible to create perfectly equivalent motivations for the different groups. 
The sample of artists may also seem quite small in absolute terms, but in practice convincing 50 busy professionals to lend their time was far from trivial. 
Our stimuli range attempted to cover a range of artistic expressions, but only represents a fraction of what is considered art across different cultures, and is biased by being produced by a single artist. However, limiting the whole experiment to 8 trials helped with keeping it short and fun, minimizing the risk of dropouts. In fact, many participants, artists and laypeople alike, reported that the experiment was exceptionally entertaining (e.g. compared to other studies), especially regarding the creative task. Our results should be interpreted with these trade-offs between generalizability and practicality in mind.

As a further consequence of these trade-offs, our models yielded effects that, while numerically modest in terms of embedding distance, were statistically robust and in line with our preregistered predictions for the central first four hypotheses. We can view the tasks as a noisy environment, where information gets lost when i) participants translate their ideas into prompts, ii) generative AI translates those prompts into images, and iii) the embeddings translate the images into quantifiable distances. In this light, the fact that we still find meaningful distances after these various transformations can be seen as a strong result. This is corroborated by the fact that artists were not necessarily working within their comfort zone, but within an artificial task that was constrained for the sake of experimental control. Yet even under these constraints, they outperformed laypeople, which further supports a deeper underlying difference between the groups. Moreover, it has been argued that it is namely constraints that give rise to creativity and novelty \parencite{feiten_constructive_2023}. In that sense, our task might have shown that it is particularly suitable, despite the limitations outlined above.

In general, our study provides a methodological framework that allows for quantifying the use of novel generative technologies via direct behavioral experimentation. Besides the visual domain, it would be informative to carry out similar experiments with artistic professions in performance, film or music, other creative and content-producing professions, or people now identifying as "prompt engineers", using various generative models relevant to their disciplines. One could assess how much of an edge professional background provides in concrete tasks, compared to an untrained person using AI assistance, or a person completing an equivalent task without any AI tools. 

We also carried out a comparison to an LLM, prompted to complete the same task, for which we had no a priori prediction. Surprisingly, it performed on par, if not better, than the average artist. Recent research has carried out similar comparisons of humans and machines, although often using crowdsourced laypeople to represent the former \parencite{begus_experimental_2024, haase_artificial_2023, koivisto_best_2023,porter_ai-generated_2024}. However, a more pressing question with economic and social implications concerns professionals, as various fields may need to reevaluate aspects of training and daily practices. We show that behavioral experiments can be used to inform these processes. 

GPT-4o performance in the creative task does not necessarily mean that "AI is more creative than humans now". The top results were still all human. The LLM-generated prompts were longer on average and quite descriptive, leading to images that in the creative task  diverged more in CLIP embedding distance from the reference on average (but we excluded that this was due to a large number of random words; see SI). The AI system was good at optimizing on our key measure of distance, but part of the reason for this might be the constrained nature of the task.  
The perhaps interesting takeaway here is that many professional artists with years of training and experience, not to mention laypeople, were \emph{not} able to score much better in this (admittedly narrow) task than a simply next-word-predicting large language model.

Images generated from prompts written by GPT-4o were also less diverse than the images produced by artists, indicating potential issues at this different level --- although one could certainly build the pipeline in a different way to address this. We used only a single instruction prompt per task, so the inputs differed only in terms of the input image, and in the creative task, the explanation of the fixed prefix. Using a larger array of different instruction prompts could potentially increase output variance as well (e.g. by prompting for various types of artists, different styles, and approaches to undertaking such tasks). Future research could thus explore how images produced by independent AI agent pipelines would compare to human-prompted AI images in terms of creativity, novelty, or aesthetics. This also highlights how this field of research should be seen as highly dynamic, with models and modalities changing quickly. Researchers are actively working on making lay user experience more intuitive \parencite{yakura_iteratta_2023}, which might lower the barriers in the future. However, it is also completely possible that this development will occur orthogonally to the skill transfer from experts to generative AI.

As AI becomes more integrated into the creative sectors, tools, and workflows, the skills defining professional artists and illustrators are changing. Traditional skills remain important, but integrating them with technological tools can enhance artistic training and professional work \parencite{fathoni_leveraging_2023, pavlik_art_2024,saez-velasco_analysing_2024}. \textcite{gu_exploring_2024} show that AI tools enhance students' confidence in the creative process and suggest providing prompt engineering training to aid creativity and cognition. Generative AI may thus unlock new possibilities for artistic expression, which includes empowering individuals to realize visions that were previously unattainable due to physical or skill limitations. Contrasting with artificial intelligence on its own, this potential has recently been referred to as "collaborative intelligence" \parencite{mollick_co-intelligence_2024} or "generative synesthesia" \parencite{zhou_generative_2024}.
Professional artists and amateurs may also use AI tools simply for different ends, the former for concrete work related tasks, or to try new styles, the latter for e.g. entertainment and exploration \parencite{elfa_using_2023,braguez_ai_2023, guljajeva_artist-guided_2024,shen_measuring_2023,tang_exploring_2024}.

These developments are of course not without their problems and challenges. Concerns have been raised about the legal and moral issues around AI training data \parencite{goetze_ai_2024}, the ethics of artistic AI usage and its impact on the creative professions and labor markets \parencite{lima_public_2025, lovato_foregrounding_2024,miyazaki_public_2024,walkowiak_generative_2024}. 
Nevertheless, it is clear that generative AI enables prototyping, content creation and curation at a groundbreaking pace and scale. This does not only affect text and visual art anymore, as video and music generation tools are beginning to mature, and cultural content is increasingly becoming generated or mediated by machines \parencite{brinkmann_machine_2023}. Disruptions in the arts are not unprecedented, however. Photography and film redefined artistic expression over a century ago, but oil painting remained relevant, itself having revolutionized art half a millennium earlier.
Education should be critically engaging with the possibilities and limitations of such tools, as well as their underlying technical principles and ethical considerations, just as with other tools and materials.

\section{Conclusions}

We carried out a controlled behavioral experiment to compare visual generative AI tool usage by artists and untrained laypeople. Unlike preceding similar research, we recruited a sample of active professional artists as the domain experts. In our two tasks, we found that expertise leads to better results in both 
accurate
copying and implementing divergent ideas.
Then again, the laypeople's results were only a small step behind.
    
We also experimented with letting an example AI language model, GPT-4o, complete analogous tasks. We found it performs on par, if not better than humans in some cases, but does not surpass the best human results.
This study serves as a preliminary exploration into how experts and non-experts compare in their ability to utilize novel and transformative AI tools, and how AI agents themselves can solve tasks that until only recently were thought the domain of human-only proficiency.

\hrulefill

\subsection*{Author Contributions and Funding}

 T.F. Eisenmann designed the study, implemented and carried out the human experiments, wrote the preregistration, preprocessed the data, wrote the paper, and provided major edits.
 A. Karjus designed the study, analyzed the data, carried out the LLM experiments, designed the figures, and wrote the paper.
 M. Canet Sola contributed to designing the study, created the stimuli, gathered the expert participants, wrote parts of the paper (mostly Method), produced the online dashboard, and co-designed the conceptual figure. 
 L. Brinkmann designed the experimental infrastructure and co-designed the conceptual figure. 
 B.I. Supriyatno implemented and monitored the text-to-image inference backend and set up the database.
 I. Rahwan commented on the framing of the manuscript and on the figures. \\
 A.K. and M.C.S were supported during the initial research process by the CUDAN ERA Chair project for Cultural Data Analytics, funded through the European Union Horizon 2020 research and innovation program (Project No. 810961). The funders had no role in study design, data collection and analysis, decision to publish, or preparation of the manuscript.

\subsection*{Acknowledgments}

 We would like to thank Yi-Tong Chen, Samira Fakhri, Diana Paola Americano Guerrero, Valerii Chirkov, and Omar Sherif for their assistance with implementing the experimental design. We further thank Ivan Soraperra for discussion on the statistical analyses and Yvonne Bialek for her help with reaching out to professional artists.

\subsection*{Data Availability}

{\raggedright
The code and metadata to reproduce the analyses is available at 
\url{https://github.com/andreskarjus/genAIexperiment}. 
The preregistration document is available at \url{https://aspredicted.org/t3mz-yjpz.pdf}.
The image data is available at 
 \\ \url{https://doi.org/10.5281/zenodo.14710706}
. See also 
 \url{https://artistlaypeopleaiexperiment.github.io} 
for an interactive dashboard to explore the data. 
\par}

\subsection*{Ethics approval and consent}

We received ethical approval for the study beforehand from the IRB of the Max Planck Institute for Human Development, approval number A2024-15. Participants gave informed consent to participate in the study before their data was collected.

\subsection*{Competing interests}

The authors have no relevant financial or non-financial interests to disclose.

\FloatBarrier
\printbibliography

\newpage
\FloatBarrier

\section*{\fontfamily{phv}\Large\selectfont Appendix}

\hrulefill

\renewcommand{\thefigure}{S\arabic{figure}}
\setcounter{figure}{0} %
\renewcommand{\thesection}{S \arabic{section}}  
\setcounter{section}{0} %

\section*{Further Explorations: How do People (and Machines) Use an Image Generator?}

\begin{figure}
	\noindent
	\includegraphics[width=\columnwidth]{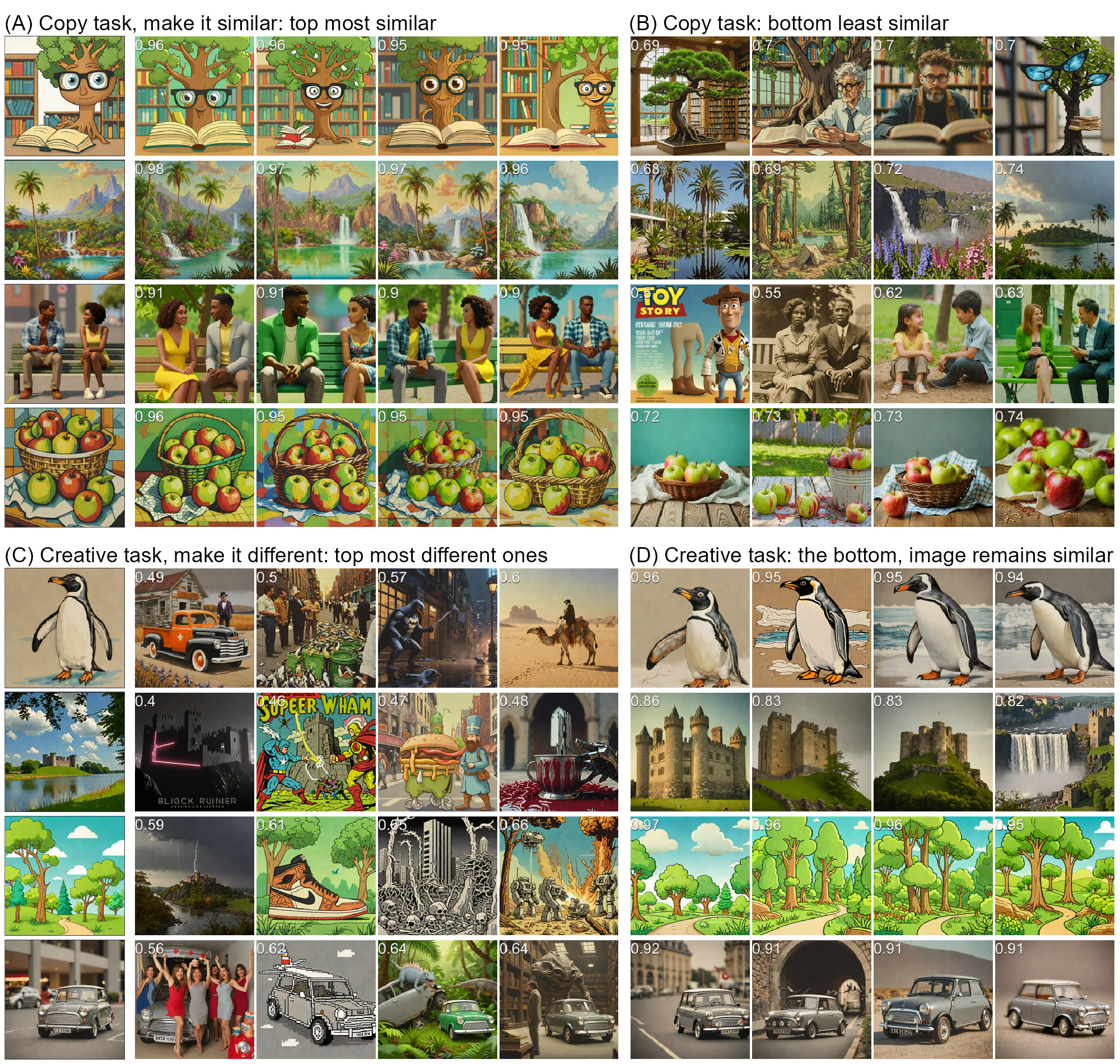}
	\caption{Examples of most and least successful trials by human participants across the experiment; cosine similarity in the corners. The reference images are in the leftmost column. This supplements the figure in the main text, providing not only the best but also the worst attempts for comparison. 
    }\label{figure_examplegrid_full}
\end{figure}

\begin{figure}[htb]
	\noindent
	\includegraphics[width=\columnwidth]{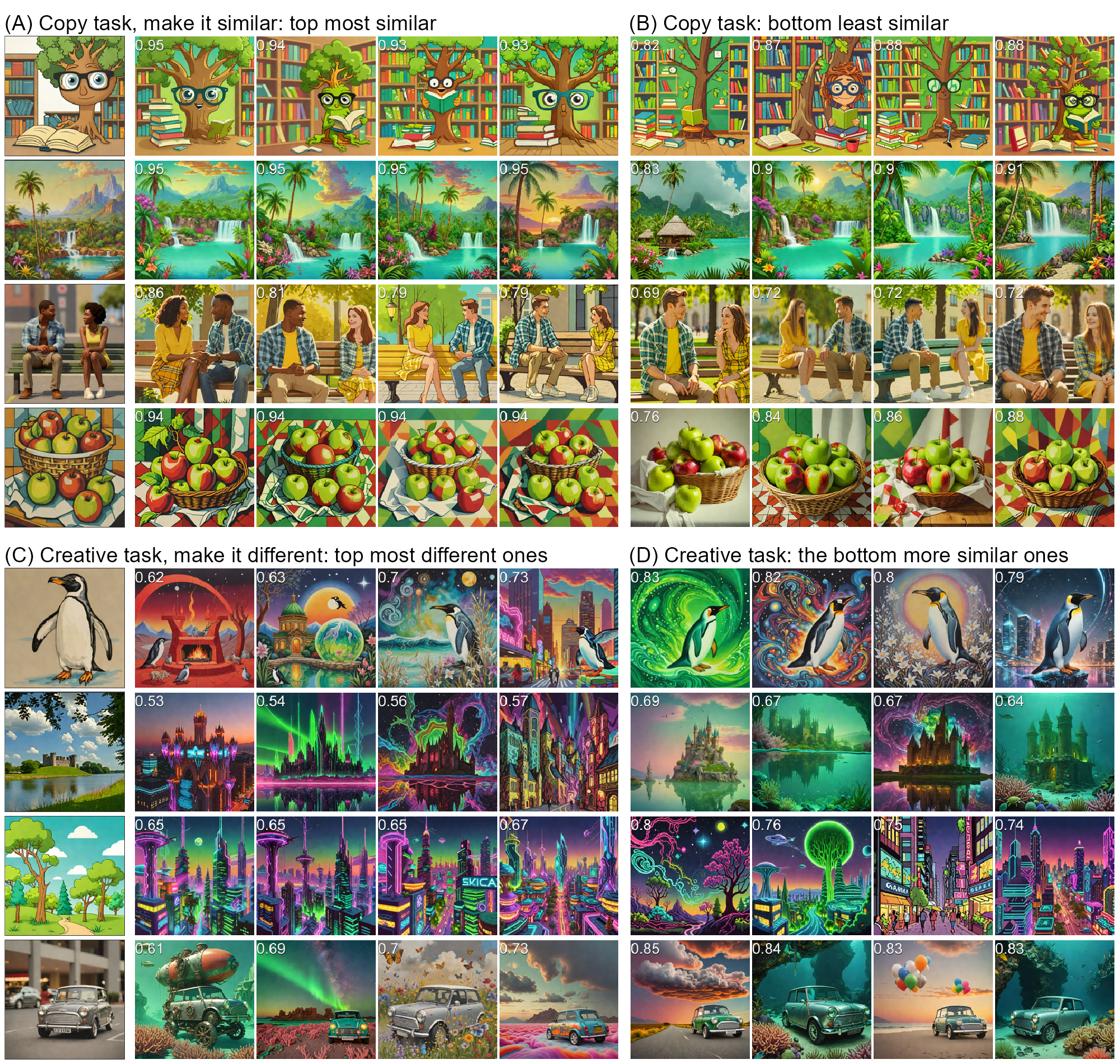}
	\caption{Examples of most and least successful trials by GPT-4o; cosine similarity in the corners. The LLM was prompted to complete the same trials in a comparable setting, and performed on par (or even slightly better) than humans. GPT's worst attempts are often better than the worst of human participants, demonstrating good adherence to instructions. In the creative task, the prevalent strategy seems to have been to overwhelm the image with new objects and colors. The reference images are again in the leftmost column.
    }\label{figure_examplegrid_gpt}
\end{figure}

This section serves to further illustrate the variable results produced by our human participants as well as the semi-autonomous image generation agent, and contains a number of supplementary figures. Figure \ref{figure_example_one} depicts the reference images and all results from one example participant to illustrate the (little) variability between the variants, as well as the solutions to the task by one creative individual.  Figures \ref{figure_examplegrid_full} and \ref{figure_examplegrid_gpt} expand the top examples graph in the main text, showing the best and worst attempts by both our human participants as well as those in the additional LLM experiment.

For the copying task, precise and relevant wording naturally worked the best, e.g. the top result for the jungle image was prompted as "a painting of an idyllic landscape, lake, mountains, waterfalls, palm trees, matte painting, detailed, tropical, jungle, inspired by mark keathley". Some humorous prompts still worked fairly well, for example, "a bad Cezanne style painting of green and red apples in a baskek with ceramic tiles behind it" resulted in an average 0.93 similarity (the typo may have affected the outcome). 
While the copying task had participants mostly just trying to describe the scenes as best they could, the creative task elicited various interesting strategies.

Some participants either knew or realized that text to image generators can be confused by compounds. Both of these prompts made the output diverge from the reference: "\textit{Photograph of a medieval castle} caricature walking down a busy new york street eating a hotdog. pastel colours" (third best castle result), and "\textit{Painting of a penguin} colored pickup truck in rural America. The truck is black, white, and orange, but the driver is dressed up as the Joker. Photorealistic" (the best penguin result; the locked prefixes in italics). This did not always work, however, e.g. in "\textit{A cartoon landscape with trees} bark, texture picture". Making use of the word allowance to describe a very different style or objects often worked: "\textit{Photograph of a medieval castle} in black and white, neon lights, blade runner, 4k unreal engine, octane. Night-time blurred motion lights" (best castle result) or "\textit{A photo of a gray classic Mini Cooper} is in the closet of Maria. She wears a red dress and her 20 friends are with her celebrating her birthday." (top car result). An attempt to hide the penguin in a box also worked quite well, as did an idea to transform the castle into a tattoo (for these further examples, see the SI).

In terms of purely quantitative image embedding similarity, the differences between human and GPT performance on the tasks were rather small, on average. However, based on close reading of the prompts and images, humans seem to have utilized more variable underlying strategies in the creative task. Given the currently utilized single instruction input (see SI), GPT-4o mostly produced prompts that led to overwhelming the output with other objects or different scenery, e.g. "\textit{A cartoon landscape with trees} transforms into futuristic cityscape, towering skyscrapers with neon lights, bustling drones flying, abstract robotic figures gliding past, shimmering holographic displays projecting vibrant surreal content." A common (but successful) theme was placing the objects underwater, e.g. "\textit{A photo of a gray classic Mini Cooper} transformed into a steampunk airship floating in a vibrant underwater fantasy world, spires of colossal coral retro-futuristically outfitted with gears gleam luminously". It was instructed to be expressive and detailed, which paid off in the copying task as well, where precise prompts led to faithful copies of the reference image, e.g. "Tropical landscape with lush greenery, vibrant flowers, tall palm trees, cascading waterfall, serene turquoise river, and majestic mountain backdrop under a warm, colorful sky. Bright, vivid colors, serene atmosphere." or "Tree character with glasses reading a book, surrounded by bookshelves, colorful books, cartoon style, whimsical and educational setting, vibrant colors, playful and imaginative atmosphere." Given these reference images and instruction prompts, GPT-4o did use the word "vibrant" a lot: this was the top non-function word it used (in 48 outputs of the 80). The LLM was instructed to make use of the word allowance, and as a result its outputs were quite long, on average 201 characters across all trials, compared to both laypeople (103) and artists (116).

\begin{figure}
	\noindent
	\includegraphics[width=\columnwidth]{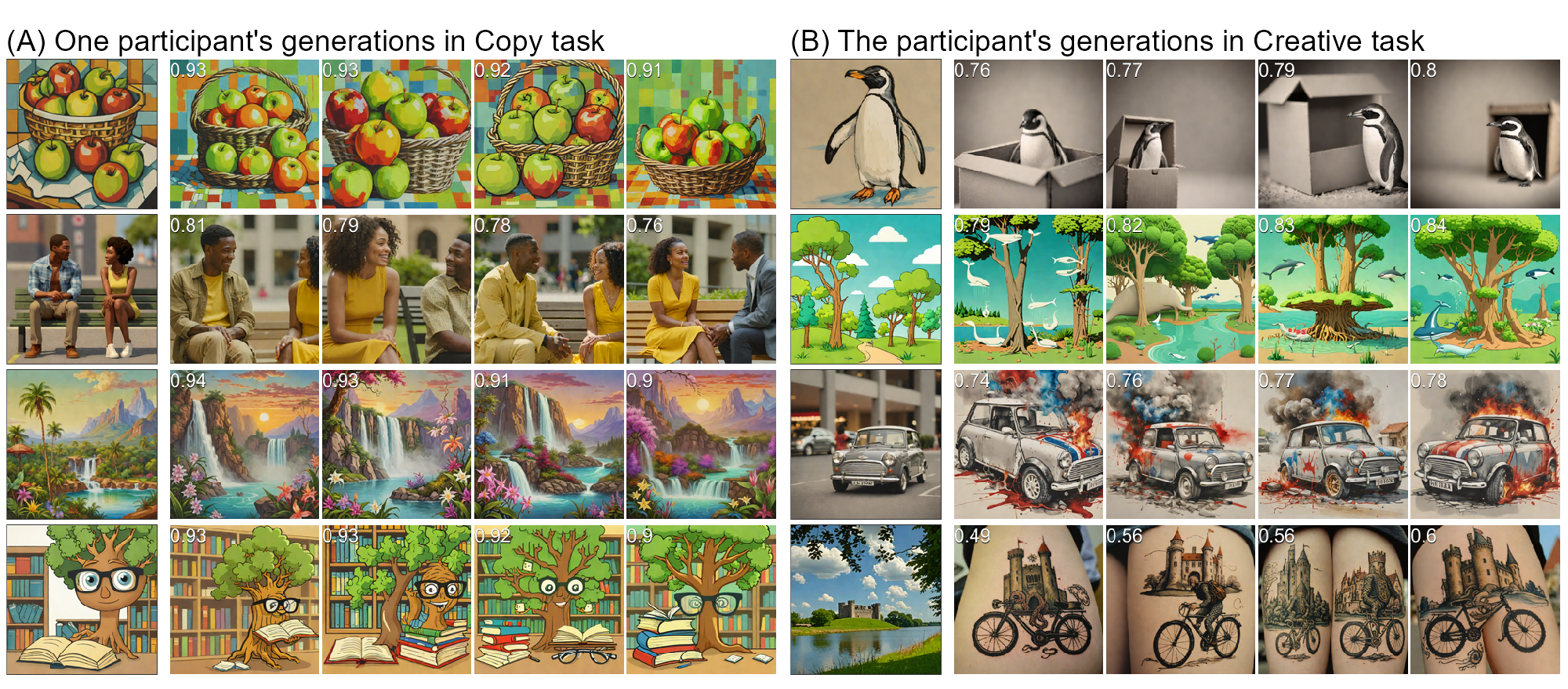}
	\caption{All creations by one artist participant. The column on the left are the reference images; cosine similarity in the corners of the results. The participant used the following prompts for the copying task (A): "abstract painting of green and red apples in a basket. Orange and grayish blue checkered background", "two persons of color sitting on a bench talking. man sitting to the left. female in yellow dress to the right. blurred builing in the background", "fantasy painting of waterfall. exotic flowers in the foreground. mountains in the back. sunset and a few clouds", "cartoon tree with glasses reading a book at a library. Cartoon drawing". The creative task (B) - with the fixed prefix of the prompt highlighted in italics here: "\textit{Painting of a penguin} in a box hiding. black and white photo motionblurred", "\textit{A cartoon landscape with trees} negative and upside down. skulpture and performed by whales wearing hats", "\textit{A photo of a gray classic Mini Cooper} exploded by a carbomb. watercolor with red and blue colors only. painted by a monkey", and "\textit{Photograph of a medieval castle}, small tattoo on a octopus wearing pants and on a bycycle". In the latter, the participant managed to refocus the generator's attention, successfully backgrounding the original concept.  %
    }\label{figure_example_one}
\end{figure}

\FloatBarrier

\section*{Additional Insights into the Experimental Data}

\begin{figure}[ht]
	\noindent
	\includegraphics[width=\columnwidth]{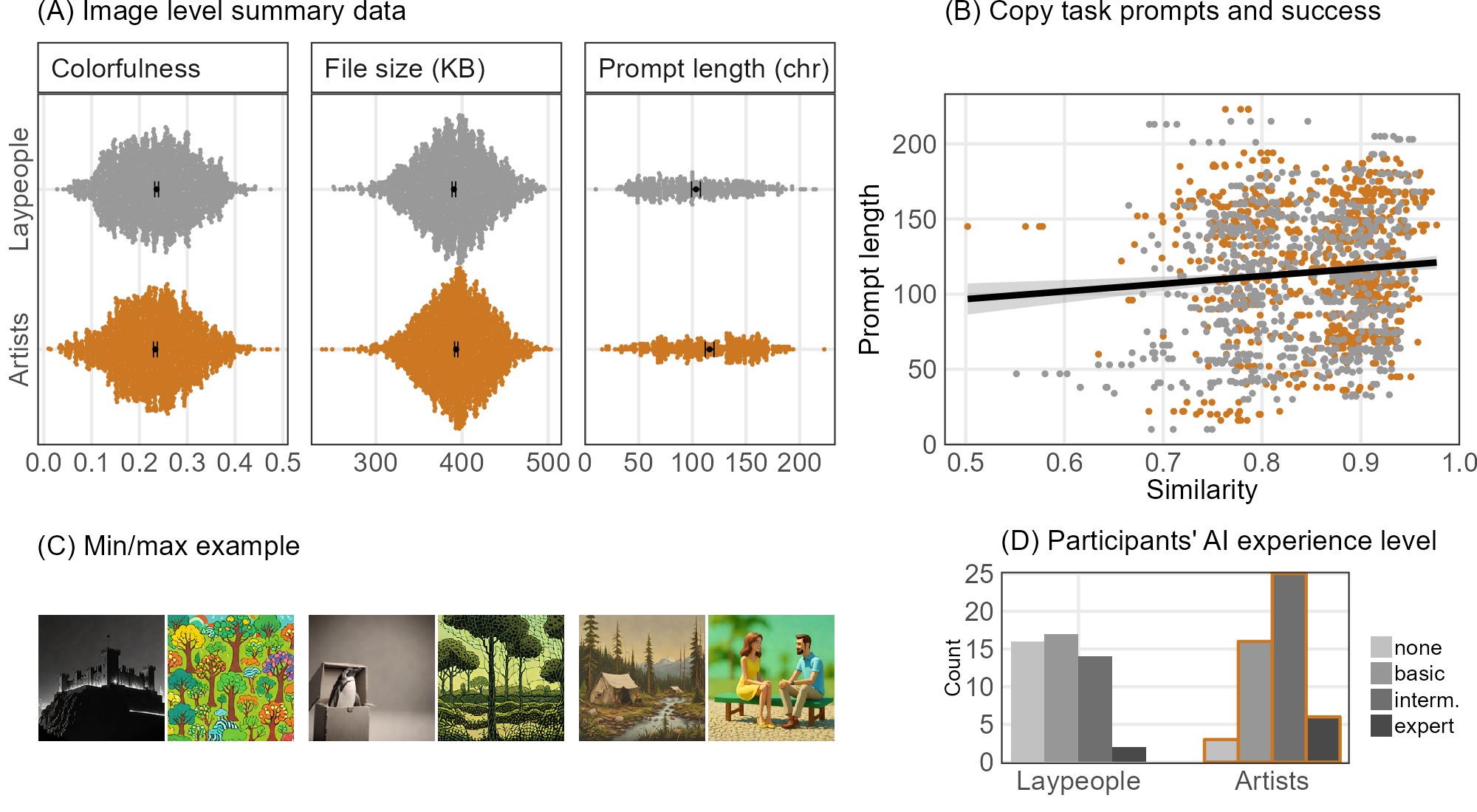}
	\caption{
    Overview of the experiment data, including all variant images. Panel (A): image colorfulness, estimated complexity (compressed size) and prompt length (characters) are fairly similar for both groups (each point is an image). There are fewer data points in the latter panel, as the prompt is the same for all four variants. Panel (C) displays the image with the lowest and highest value in each of the metrics, e.g. for colorfulness (left) the gray-scale castle is the lowest while the bright cartoony forest is the most colorful. The complexity (middle) of the monochrome penguin with a plain background is lowest while the intricately textured green forest is highest. Longer prompts can help specify an output, but shorter prompts can still create complex images, e.g. the cabin (right) is generated by a one-word "wilderness". Longer prompts can help but do not always lead to more accurate depictions when copying (B). Panel (D) illustrates prior experience with AI tools on a scale of 0 (none) to 3 (expert).
    }\label{figure_summaries}
\end{figure}

While not the main object of study, we also measured several summary statistics to gain some insight into the data (Fig. \ref{figure_summaries}): image colorfulness, complexity, and prompt length (in characters). For color, we used the "M3" measure from \textcite{hasler_measuring_2003}, and for complexity the file size of the PNG compressed image. Compression is a well-known estimate of image complexity that aligns with human perceptions \parencite{machado_computerized_2015,chamorro-posada_simple_2016,karjus_compression_2023,muller_compression_2018}.
In general, we find no notable differences in terms of image colorfulness or complexity between groups.
The only slight difference in average prompt length between groups was not significant (simple linear regression, $\mathrm{F}(1, 3146) = 3.56, p = 0.06$).

\section*{Adding Random Words to the Prompt Does Not Help Much on Average in the Creative Task}

As discussed in Methods, we also experimented with two random baselines to make sure the creative task could not be completed successfully by just adding random words or gibberish to the prefixed part. Neither leads to better results than the human (or GPT) written prompts, on average, but a long list of random words can indeed do better than a badly written or too short of a prompt, when the task is to write a prompt that would diverge from the fixed prefix as far as possible. 

\begin{figure}[htb]
	\noindent
    \includegraphics[width=0.6\columnwidth]{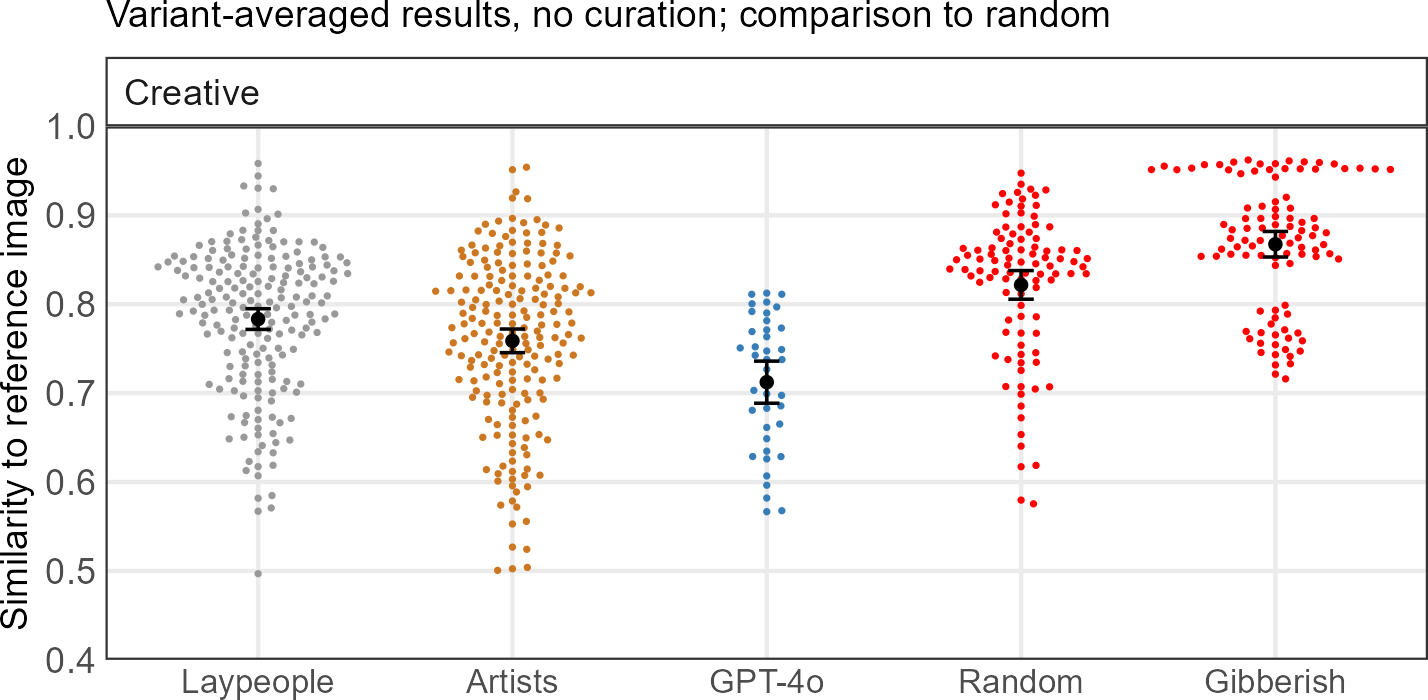}
	\caption{
          Experiment results for the creative task, where the goal was to create images with a low similarity to the reference (variant-averaged, no curation), compared to two random baselines: prompts consisting of randomly sampled English words, and of randomly constructed gibberish pseudo-words. 95\% confidence intervals (black) added for reference.
    }\label{figure_random}
\end{figure}

\FloatBarrier

\newpage   %
\section*{Human and GPT Results Trial by Trial}

\begin{figure}[ht]
	\noindent
	\includegraphics[width=\columnwidth]{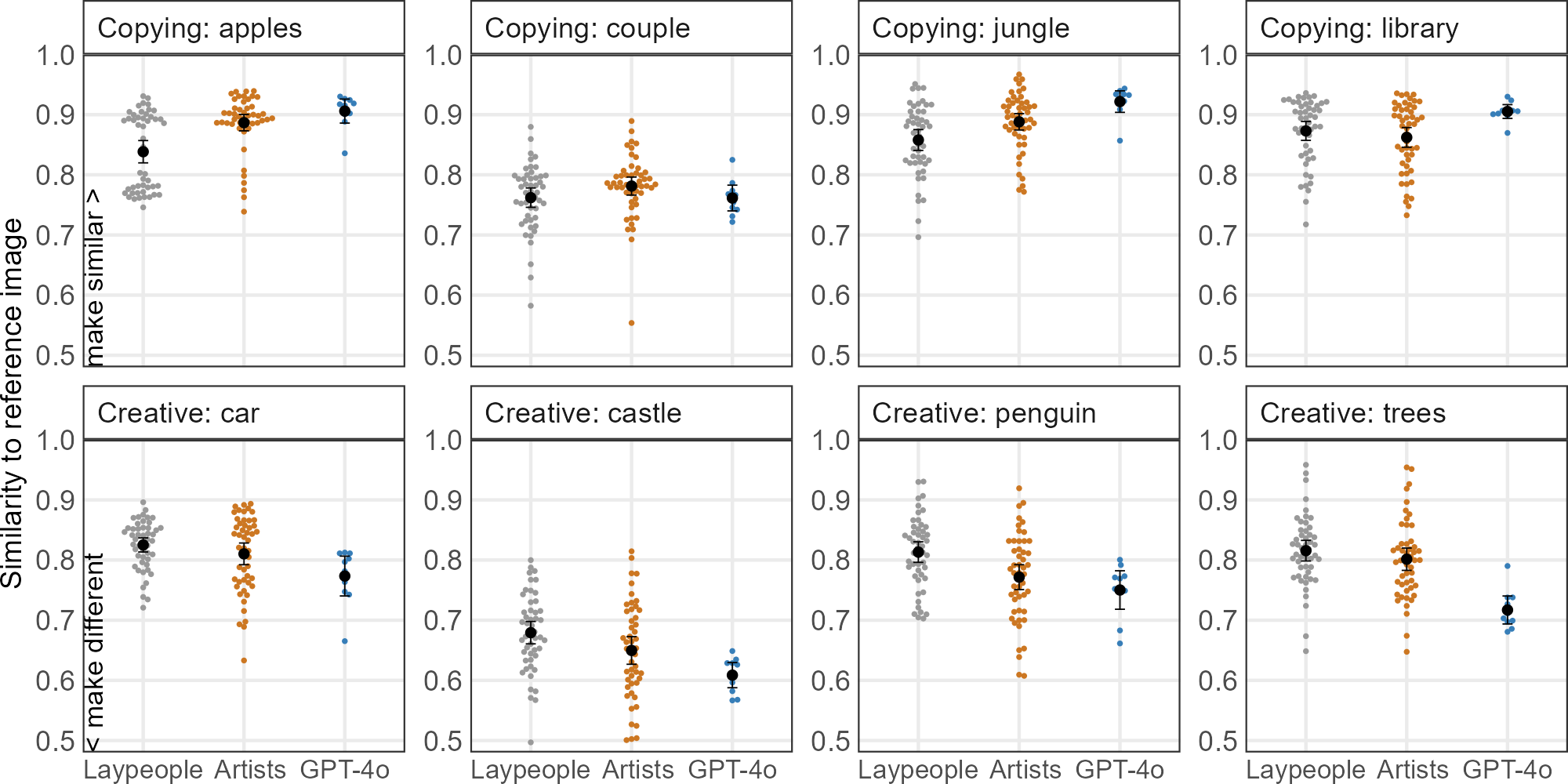}
	\caption{Experimental results by trial. As in the figure in the main text, each point is the averaged cosine similarity of the generated image quartet to the reference image, and each point represents the attempt of one participant. 
    }\label{figure_trial_comparisons}
\end{figure}

Figure \ref{figure_trial_comparisons} expands the main results figure of the main text, showing results for each trial separately (without curation). While there is variation between the trials (which we control for in statistical modeling), the overall trends remain the same. Some trials were more difficult than others. Creating a copy of the apples still life painting was comparatively easier than copying the couple on the bench, but the former proved challenging still for about half the laypeople (notice the bimodal distribution). Making the castle or penguin different from the reference was easier than doing the same with the Mini Cooper.

\vspace{1cm}

\section*{Stimuli Generation Prompts}

The copying task reference image were created with Stable Diffusion using the following prompts. See the figures above for the images themselves, here referenced to using single-word shorthands: \\
Library: cartoon style tree character with black glasses and big blue eyes and a book in the left in a library  \\
Jungle: oil painting of a colourful African landscape with a river, a jungle with palm trees, flowers, a waterfall and a mountain on the horizon \\
Couple: A 3D-rendered game screenshot featuring an African American man on the left and a woman on the right, sitting on a bench at a bus stop. They are facing each other in a friendly manner, engaged in conversation, set against an urban street backdrop.\\
Apples: An oil painting of a full basket of apples in the Cubist style, inspired by Picasso

The creative task reference image were created with Stable Diffusion using the following prompts; the fixed parts that the participants could see but not edit are shown in bold: \\
Penguin: \textbf{Painting of a penguin} in a student-like style \\
Castle: \textbf{Photograph of a medieval castle} with four watch towers centered by a  lake, framed by lush grass and scattered trees. The castle stands prominently under a bright blue sky, dotted with impressive cumulonimbus clouds. \\
Trees: \textbf{A cartoon landscape with trees} and blue sky \\
Car: \textbf{A photo of a classic gray Mini Cooper} in a parking lot of a shopping mall

\section*{Prompts Used in the LLM Experiment}

The prompts used to instruct GPT-4o to complete the experiment were as follows. The reference image, mentioned in the prompt, was sent to the API along with the prompt.\\
The copying task: \textit{
You are an expert at visual art, photography, drawing, painting, art history, and an expert prompter. Your task is to create an image by making use of a generative AI model, our "image machine". To do so, you need to provide text inputs, so-called "prompts". You are being shown an image. Your goal is to write a detailed prompt that would result in an image that is as close as possible to this image in all aspects, including main subjects and where they are on the image, and image composition, colors, style or genre, etc. Describe the shape, size and color of prominent objects, and relevant characteristics (gender, age, appearance, ethnicity, etc) of any focal human subjects. The prompt should be up to 30 words in length, so be concise. Do not comment, just provide the 30-word prompt without quotes.
}

The creative task: \textit{
You are a creative visual artist and an expert prompter. Your task is to create an image by making use of a generative AI model, our "image machine". To do so, you need to provide text inputs, so-called "prompts". You are being shown an image. Your goal is to write a prompt that will make a new image as different from the original as possible. Pro tip: Ideally, you should try to aim for creating an image with different content as well as different visual style, as unrelated to the original as possible. Also, negation does not work very well here - if you state "an image with no elephants", you are actually more likely to get an image with elephants! So do not use negation in the prompt.\\
Now look at this image. Try to write a prompt that would create an image that is as different as possible from this image. The prompt should be up to 30 words in length in total but not longer. However, you are constrained in that your prompt MUST start with this phrase describing the original image:\\
/the fixed lead of the given trial, e.g. "A cartoon landscape with trees"/ \\
Start with that phrase, but then try to write the rest of the prompt so that its resulting image would actually not contain or would obscure anything mentioned in that phrase. Use clever wordplay and think outside the box to get away from the meaning and appearance of the reference image and its descriptive phrase. Your prompt, while containing this initial phrase, should describe a new image that would look as different as possible from the reference image you are seeing right now, and would be nothing like /the fixed lead/. Do not comment, just provide the 30-word prompt without quotes.
}

\section*{Prompts Used in Zero-Shot Analytics}

The zero-shot instructions for the prompt components feature analysis were the following:\\
\textit{Analyze this short quoted Text describing an image and output a comma-separated list of elements that are mentioned. Output only these keywords where relevant, as explained and exemplified (in brackets) here. These are the only valid output terms:\\
style = if Text mentions artistic style or period (like cubist, expressionism) \\
genre = mentions genre (like landscape, still life, fantasy)\\
medium = mentions type or medium of image (like photo, painting, cartoon)  \\
colors = mentions any colors (including phrases like colorful or bright colors)\\
adjectives = contains any adjectives or similar descriptives (like pleasant, misty, dystopian, destroyed)\\
objects = mentions one or more tangible subjects or objects (like car, trees, buildings, people)  \\
setting = mentions a setting or background (like library, sea, jungle, wilderness, night)  
action = if Text mentions some action (like reading, sit, standing).\\
How to respond: only a comma-separated keywords from the above if relevant, but do not describe the actual colors or styles, and do not comment! If no none of the above a present, just say None. Here is the Text to analyze:\\
""" /prompt text/ """"  }

This prompt was used to infer sentiment:\\
\textit{Below is a quoted Feedback from an experiment where people used gen AI to create images. We asked them how they feel about AI. Summarize the Feedback sentiment as one of these categories that best matches its primary sentiment:\\
Positive = mostly postitive, excited, praising or enthusiastic attitude towards AI.\\
Critical = a critical, concerned or dissatisfied attitude or talks about flaws or worries or implies AI is not good for art.\\
Mixed = if Feedback expresses mixed or ambivalent feelings or mentions both positive and negative aspects (eg "fun but boring").\\
Neutral = pragmatic or indifferent Feedback, or neutral discussion about technical aspects without sentiment or emotion (eg "task was challenging" - this is not negative about AI as such discusses the task they completed beforehand).\\
Output a single category, do not comment! This is the Feedback:\\
"""Opinion on generative AI: /the opinion/ """ }

\end{document}